\documentclass[3p,preprint]{elsarticle} 
\usepackage{epsfig}
\usepackage{amsfonts}
\usepackage{amsmath}
\usepackage{amssymb}
\usepackage{xcolor}
\usepackage[normalem]{ulem}
\usepackage{ulem}

\usepackage[titletoc,toc,title]{appendix}

\usepackage[bookmarks=true]{hyperref}

\begin{document}

\title{Characterization of helical states in semiconductor quantum wells using quantum information quantities}

\author{Natalia Giovenale}
\ead{ngiovenale@famaf.unc.edu.ar}
\author{Omar Osenda}%
 \ead{osenda@famaf.unc.edu.ar}
\address{Facultad de Matem\'atica, Astronom\'{\i}a, F\'{\i}sica y 
Computaci\'on, Universidad Nacional de C\'ordoba e Instituto de F\'{\i}sica 
Enrique Gaviola - CONICET,  Av. Medina Allende s/n, Ciudad Universitaria, 
CP:X5000HUA C\'ordoba, Argentina,
}%


\date{\today}

\begin{abstract}
The information content of   one-electron bulk and edge states 
in semiconductor quantum wells is calculated in the inverted regime, where 
edge states, topologically protected,  are responsible for the conduction in Spin Quantum Hall 
effect 
experiments. To study the information content of these states we first calculate 
realistic two dimensional one-electron states, solving first the eight-band 
$\mathbf{k}\cdot\mathbf{p}$  Hamiltonian to obtain the bulk states and then a 
four band effective Hamiltonian to obtain the edge states. The behavior of information-like quantities, as a function of the different 
parameters that define 
the quantum well, is analyzed. The results presented show that the helical edge states can be singled out using different quantities that characterize the rich phenomenology of these states.
\end{abstract}

\maketitle


\section{\label{sec:introduction}Introduction}

Semiconductors, and the nanostructures made of them, allow to investigate an 
incredible set of physical situations because, among other reasons, changing 
the density of charge carriers by several orders of magnitude is a routine task 
nowadays, and this leads to remarkably different scenarios. Since the experiments that 
led to the discovery of the 
Quantum Hall effect by  Klaus von Klitzing \cite{vonKlitzing1980}, the 
semiconductor quantum well has 
become an ideal scenario to study a host of different phenomena: fractional 
Quantum Hall effect \cite{Laughlin1983},  Quantum Spin Hall effect 
\cite{Kane2005, Bernevig2006,Koning2007} and condensation of excitons and 
polaritons 
\cite{Snoke2002, Wang2019}, just to name a few.

In the case of the Quantum Spin Hall effect, the properties of the helical edge 
states, which are topologically nontrivial \cite{Bernevig2006}, are fundamental to understand the conductance properties in HgTe quantum 
wells. 
The characterization of topological states can be done using at least two 
approaches: calculating a topological quantity, as the Chern number or the $Z_2$ 
number \cite{Kane2005,Yu2011,Schnyder2008,deLisle2014}, or calculating the 
entanglement entropy or entanglement spectrum of the 
state of interest \cite{Kitaev2006,Li2008,Sterdyniak2012}. The first approach mentioned
can be implemented when the band structure is at disposal, while the second 
approach has been mostly used  to study many-body models when the 
reduced density matrix of a subsystem, that is half the whole system, can be 
effectively calculated.

The entanglement 
entropy and the entanglement spectrum capture the non-local character of a 
topological state, and have been calculated for spin chain states 
\cite{Isakov2011,Jiang2012,Grover2013,Lee2013}
and for states that are though to be good approximations of the many-electron 
ground state 
 wave function of the fractional Quantum Hall effect, {\em i.e.} 
the Laughlin states \cite{Haque2007,Li2008,Zozulya2007,Zozulya2009}. Other studies 
have analyzed  the general properties of the entanglement spectrum of 
topological 
states in different physical setting and geometries 
\cite{Lauchli2010,Fidkowski2010,Poilblanc2010,Hsieh2014,Zaletel2013,Cincio2013}, and it is been used to reveal the presence of a Haldane phase 
\cite{Pollman2010,Lisandrini2017}. All in all, depending on the physical systems, sometimes the 
topological character is better put into evidence using the behaviour of an 
entanglement entropy, while in other situations the best tool is the 
entanglement 
spectrum.

Since the properties of electrons in semiconductor quantum wells are well 
described using the $\mathbf{k}\cdot\mathbf{p}$ model, it is quite natural to 
ask if the one-electron edge states obtained in this way can be analyzed using some of the 
techniques developed to study topological quantum states in quantum spin 
chains. There are 
two main difficulties to proceed along this idea. The first one has to do with 
the difficulties involved in obtaining realistic one-electron edge states. If a one-electron edge state 
is provided, then  the second issue must be addressed: Which is the subsystem 
that  is to be traced out to obtain the necessary reduced density matrix to 
calculate entanglement entropies and spectrum?

Realistic  one-electron edge states in semiconductor quantum wells are quite 
difficult to calculate \cite{Scharf2012}, in contradistinction with the 
spectrum, which can be 
obtained trough well educated guesses about the envelope functions that are the 
solutions of the eigenvalue problem for some multi-band Hamiltonian. The reasons 
behind this difference can be easily traced. Edge states depend on, at least, 
two coordinates, there are matching and contour conditions to be satisfied and 
the physical regimes of interest arise when there are a few bands that 
interact, making the calculation in terms of multi-band Hamiltonians mandatory. 
Recently,  a number of works have addressed this issue and paved the way to 
obtaining robust edge states with and without an external magnetic field applied 
to the quantum well 
\cite{Krishtopenko2018,Krishtopenko2016,Skolanski2018,Chen2019}. The second difficulty mentioned above can be dealt with by using real-space 
entanglement \cite{Sterdyniak2012}, in which the subsystem that is traced out 
is a finite domain of one (or 
several) of the spatial coordinates of a given quantum state.

  The aim of 
this paper is to study the properties of one-electron edge states confined in HgTe-CdTe quantum wells. This system has been extensively studied since it shows the appearance of helical edge states, which govern the transport properties in the Quantum Spin Hall (QSHE) and Quantum Hall (QHE) effects. Conductance experiments show, when the appropriate external magnetic field is applied, the presence of topological states. So, for quantum wells defined with parameters close to those of the QSHE regime, we obtain the edge states and characterize them using different quantum entropies, under the hypothesis that these entropies would put into evidence physical traits as localization, polarization and others.  

The paper is organized as follows, the eight-band Kane Hamiltonian is presented 
in 
Section~\ref{sec:kane-hamiltonian}, together with the effective two-dimensional Hamiltonian. 
The details involved in the numerical calculations needed to obtain the bulk states and spectrum are deferred to \ref{ap:Kane-Hamiltonian}, while the corresponding details concerning the edge states are consigned in \ref{ap:edge-spectrum-states}.  In both cases, bulk states and edge states, the Rayleigh-Ritz variational method is employed to obtain a high-precision numerical approximation for the spectrum and eigenstates.  The study of the edge and normal states 
entropies is developed in Section~\ref{sec:entropies}, and some complementary results are included in \ref{ap:real-space-entropy}.  In this Section we also present the definition of the spatial von Neumann entropy, the corresponding reduced density matrix and the numerical quantities that enter into the implementation of the Rayleigh-Ritz method. Finally, we 
present our conclusions and a discussion of our results in 
Section~\ref{sec:conclusions}.

\section{Kane Hamiltonian, effective Hamiltonian, bulk and edge states}\label{sec:kane-hamiltonian}

\begin{figure}
    \centering
    \includegraphics[width=0.6\linewidth]{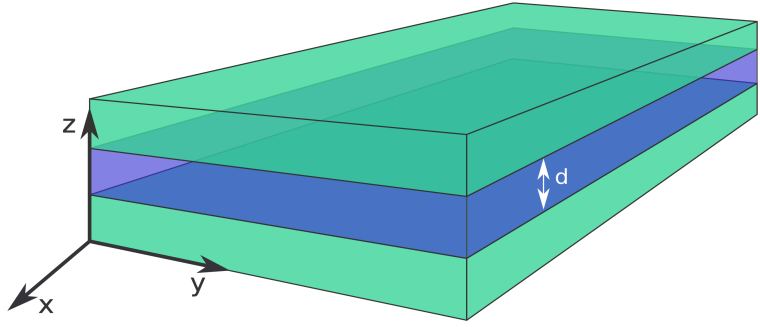}
    \caption{\label{fig:cartoon} The cartoon depicts the layered structure that 
forms the quantum well. From bottom to top, there are three layers composed of 
Cd$_{0.7}$Hg$_{0.3}$Te, HgTe and Cd$_{0.7}$Hg$_{0.3}$Te, respectively. The middle layer has a width $d$ in the $z$ 
direction, that is transversal to the three slabs. The reference coordinate 
system is also depicted in the lower left corner of the cartoon.}
\end{figure}

Quantum Wells (QW) can be described as a sandwich of three semiconductor slabs, as pictured in Figure~\ref{fig:cartoon}. In this cartoon, we represent the HgTe/Cd$_{0.7}$Hg$_{0.3}$Te QW that we study. As can be seen in this Figure, the middle slab, in blue, represents the HgTe slab, and has a width $d$ known as the QW width. The outer slabs, in green, are made of Cd$_{0.7}$Hg$_{0.3}$Te, and considered of infinite width.
There is a critical value of the QW width, $d_c$, for which the transport properties of the system change as a result of the band inversion \cite{Bernevig2006}. In HgTe/Cd$_{0.7}$Hg$_{0.3}$Te QW's, this value is $d_c\sim6.3$~nm. If $d<d_c$ the transport regime is normal and the bands are not inverted, while for $d>d_c$, the Quantum Hall regime, the transport occurs along the $(x,y)$ plane. From now on, we refer to this system as a HgTe/CdTe QW.

The band structure of semiconductor quantum wells can be obtained studying the 8-band $\mathbf{k}\cdot \mathbf{p}$ Hamiltonian, but a full numerical treatment of a two dimensional problem with this Hamiltonian is quite taxing. So, to obtain 
the edge state wave functions, it is necessary to 
make use of an effective Hamiltonian \cite{Krishtopenko2018}, which is obtained from the bulk approximated states of the infinite quantum well. In the following subsection, the procedure to obtain the edge states is roughly described, but more specific details can be found in \ref{ap:Kane-Hamiltonian} and \ref{ap:edge-spectrum-states}. 

\subsection{Eight-band Kane Hamiltonian and bulk states}
\label{sec:bulk}

If the quantum well is infinite, or more precisely, if there is not any boundary conditions at finite values of the $x$ or $y$ coordinates, the electronic band structure of the quantum well can be calculated using the  Kane Hamiltonian \cite{Novik2005}

\begin{equation}\label{eq:kp-ocho-bandas}
H = \left(\begin{array}{cccccccc}
    T &
    0 & 
    -\frac{1}{\sqrt{2}}P k_{+} &
    \sqrt{\frac{2}{3}}P k_z & 
    \frac{1}{\sqrt{6}} Pk_{-} & 
    0 &
    -\frac{1}{\sqrt{3}} P k_{z} & 
    -\frac{1}{\sqrt{3}} P k_- \\
%
%
    0 &
    T &
    0 &
    -\frac{1}{\sqrt{6}} Pk_{+} &
    \sqrt{\frac{2}{3}}P k_{z} &
    \frac{1}{\sqrt{2}}P k_{-} &
     -\frac{1}{\sqrt{3}} P k_{+}&
     \frac{1}{\sqrt{3}} P k_{z}\\
%
%
    -\frac{1}{\sqrt{2}} k_{-} P &
    0 &
    U + V &
    -\bar{S}_{-} &
    R &
    0 &
    \sqrt{\frac{1}{2}} \bar{S}_{-} &
    -\sqrt{2}R \\
%
%
    \sqrt{\frac{2}{3}} k_{z}P &
    -\frac{1}{\sqrt{6}} k_{-} P &
    - \bar{S}_{-}^{\dagger}&
    U-V &
    C &
    R &
    \sqrt{2}V &
    -\sqrt{\frac{3}{2}} \tilde{S}_{-} \\
%
%
    \frac{1}{\sqrt{6}} k_{+} P &
    \sqrt{\frac{2}{3}} k_{z}P &
    R^{\dagger} &
    C^{\dagger} &
    U- V &
    \bar{S}_{+}^{\dagger} &
    -\sqrt{\frac{3}{2}} \tilde{S}_{+} &
    -\sqrt{2} V \\
%
%
    0 &
    \frac{1}{\sqrt{2}} k_{+} P &
    0 &
    R^{\dagger} &
    \bar{S}_{+} &
    U+ V &
    \sqrt{2} R^{\dagger} &
    \frac{1}{\sqrt{2}} \bar{S}_{+} \\
%
%
    -\frac{1}{\sqrt{3}} k_{z} P &
    -\frac{1}{\sqrt{3}} k_{-} P &
    \frac{1}{\sqrt{2}} \bar{S}_{-}^{\dagger} &
    \sqrt{2} V &
    - \sqrt{\frac{3}{2}} \tilde{S}_{+}^{\dagger} &
    \sqrt{2} R &
    U - \Delta &
    C \\
%
%
    -\frac{1}{\sqrt{3}} k_{+} P  &
     \frac{1}{\sqrt{3}} k_{z} P &
    -\sqrt{2} R^{\dagger} &
    -\sqrt{\frac{3}{2}} \tilde{S}_{-}^{\dagger} &
    -\sqrt{2} V &
    \frac{1}{\sqrt{2}} \bar{S}_{+}^{\dagger} &
    C^{\dagger} &
     U -\Delta
    \end{array}\right) \,,
\end{equation}

\noindent where 
\begin{equation}\label{eq:kpm}
k_{\pm} = k_x \pm i k_y\,,
\end{equation}
\begin{equation}\label{eq:kz}
k_z = -i \partial_z .
\end{equation} 
 The diagonal terms of the Hamiltonian have a ``kinetic energy'' shape, and introduce the lowest energy of the bulk conduction band, $E_c$, and the largest energy of the bulk valence band, $E_v(z)$. For instance,  
\begin{equation}
\label{eq:kinetic-T}
T = E_{c}(z) + \frac{\hbar^2 }{2m_0} \left[
(2F+1) (k_x^2 + k_y^2) + k_z (2F+1) k_z
\right].
\end{equation}
The other entries on the Hamiltonian in Equation (\ref{eq:kp-ocho-bandas}),  and the meaning and values of the material parameters, as $m_0,F$ and $E_c$ in Equation~\ref{eq:kinetic-T}, are given in \ref{ap:Kane-Hamiltonian}.

The 
bulk band structure of the system can be obtained by solving the eigenvalue problem

\begin{equation}\label{eq:bulk-band-structure} 
    H \Psi = E_B \, \Psi , 
\end{equation}

\noindent where $E_B$ are the energy eigenvalues which form the bulk band structure, and $\Psi$ are eight-band spinors, whose components, $\Psi_i$, are the solutions to the eigenproblem with eight coupled equations of the form
\begin{equation}
\label{eq:bulk-eigenproblem}
\sum_{j=1}^8 \mathbf{H}_{i,j} \Psi_j(z) = E_B \, \Psi_i(z),
\end{equation}
\noindent for $i=1,2,..,8$, where $\mathbf{H}_{i,j}$ is the matricial one-coordinate 
differential operator that is obtained from Equation~\ref{eq:kp-ocho-bandas}, 
replacing  
$k_x$ and $k_y$ by the $x$ and $y$ components of a vector belonging to the first 
Brillouin zone of the semiconductor. The behavior of the electrons is 
effectively described by the resulting one-coordinate Hamiltonian, since the 
problem is 
translational invariant in the $x$ and $y$ directions. 

To solve the eigenproblems in Equation~\ref{eq:bulk-eigenproblem} we employ the Rayleigh-Ritz variational method. This method, in combination with the 8-band Kane Hamiltonian, has been successfully used to obtain the band structure of different semiconductor nanostructures \cite{Giovenale2022,Kishore2014,Krishtopenko2018}. 
  As a result, the bulk eigenstates for the quantum well are obtained, and can be written as
\begin{equation}
    \psi(x,y,z)=e^{ik_xx+ik_yy}\sum_{j=1}^8 \Psi_j(z)|j\rangle
\end{equation}
where $|j\rangle $ is the basis of the Kane Hamiltonian. For details about the implementation of the Rayleigh-Ritz method and the basis set used to solve the ploblem see \ref{ap:Kane-Hamiltonian}.

\begin{figure}[ht]
    \centering
    \includegraphics[height=6cm]{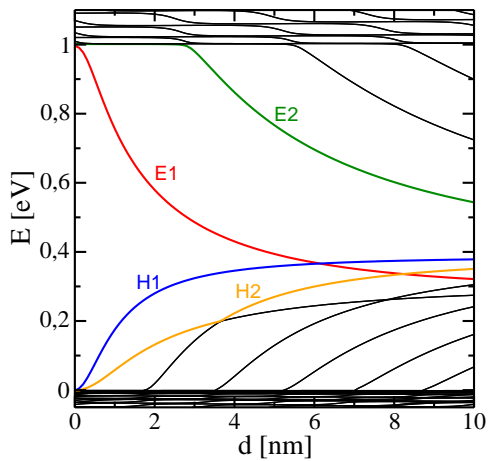}
    \caption{\label{fig:bulk-spectrum} Spectrum of the HgTe/CdTe quantum well, 
as a function of the width of the HgTe layer, $d$. 
The sub-bands $E2$ (green line), $E1$ (red line), $H1$ (blue line) and $H2$ 
(orange line)  are 
highlighted to better appreciate the eigenvalue crossing that separates both 
regimes. For $d<d_c$ the 
system is in the normal regime, $E_1 > H_1$, while for $d>d_c$ the band 
ordering is 
reverted.}
\end{figure}

Figure~\ref{fig:bulk-spectrum} shows the effect over the band structure of increasing the width $d$ of the HgTe layer. The bottom and top of the conduction and valence bulk bands, respectively, of the system can be appreciated as the black bands of energies above $1.5$~eV and below zero  (we adopt the convention that the top of the valence band corresponds to $0$~eV). It can be seen that there are energy levels that detach from both bands, and they are more numerous for increasing values of the width $d$. The convention used to name them is quite straightforward, those that come off from the conduction band are denoted as $En$, where $n$ is a natural number that corresponds to the  detachment order. The levels that become detached from the valence band are denoted by $Hn$. Figure~\ref{fig:bulk-spectrum} also shows the band inversion phenomena near $d_c\sim6.3$~nm. As can be seen, the energy of the $E1$ sub-bands falls bellow that of the $H1$ sub-band around $d=d_c$. As $d$ increases, more of this sub-band crossings occur.

\subsection{Broken symmetry Effective Hamiltonian  and edge states}

When the quantum well is finite in one direction of the $(x,y)$ plane, edge states appear under certain configurations. Following the most used convention, we consider 
a finite stripe in the $y$-direction, of length $2L=1000$~nm. To model this situation, 
homogeneous boundary 
conditions have to be imposed in  $y=-L/2$ 
and $y=L/2$ to the electron spinor components $\Psi_j$ when solving the Hamiltonian in Equation~\ref{eq:kp-ocho-bandas}. Moreover, in this configuration the quantum well has translational invariance only on 
the 
$x$ direction, so now $k_y=-i\partial_y$, and the basis set functions for solving the coupled equations in Equation~\ref{eq:bulk-eigenproblem} are functions of both $y$ and $z$, which increases considerably the dimensions of the matrix representation of the eigenproblem.

A solution to this problem was proposed in Reference~\cite{Krishtopenko2018}. Keeping in mind that near the region of interest, {\em i.e.}, for energies close to the band gap energy, the transport behaviour is dominated by the $E1, E2, 
H1$ and $H2$ sub-bands \cite{Krishtopenko2016,Krishtopenko2018}, and using perturbation theory \cite{Rothe2010,Krishtopenko2016}, the eigenstates that correspond to this sub-bands, that are obtained by the procedure described in Section~\ref{sec:bulk}, are used to generate an effective Hamiltonian of the form

\begin{equation}\label{eq:effective-Hamiltonian}
\mathbf{H}_{eff} = \left(
\begin{array}{cc}
\mathcal{H}(+k) & 0 \\
0 & \mathcal{H}^{*}(-k) 
\end{array}
\right) ,
\end{equation}

\noindent where $\mathcal{H}$ is a $4 \times 4$ differential operator in the $y$ coordinate, and $\mathcal{H}^{*}$ is its complex conjugate. The broken symmetry between the states and eigenvalues is clear, since the eigenvalues and eigenstates corresponding to positive values of $k_x$ are calculated using $\mathcal{H}$, while the ones with negative values of $k_x$ are obtained with $\mathcal{H}^{*}$.

Obtaining the eigenstates and eigenvalues of the Hamiltonians $\mathcal{H}$ and $\mathcal{H}^*$ can be done in a similar fashion as the one described in Section~\ref{sec:bulk}. Consider the associated eigenproblems 

\begin{equation}\label{eq:eigen-edgep}
 \mathcal{H}(+k)\Phi_+(y) = E_{+} \Phi_+(y) ,
\end{equation}

\noindent and 

\begin{equation}\label{eq:eigen-edgem}
 \mathcal{H}^{*}(-k)\Phi_-(y) = E_{-} \Phi_-(y) ,
\end{equation}

\noindent where $E_{\pm}$ are the eigenvalues of the band structure of the finite quantum well, $\Phi_{\pm}$ are four-band spinors and $\pm$ stands for the sign of $k_x$. 

The components of $\Phi_{\pm}$ are obtained by solving  the eigenproblem with four coupled equations for each value of $k_x$, in a similar way as in Equation~\ref{eq:bulk-eigenproblem}. As a result, the spectum of the effective Hamiltonian as a function of $k_x$ is obtained, and the edge eigenstates can be written as
\begin{equation}
    \phi(x,y,z) = e^{ik_xx}\sum_{i=1}^4\Phi_{\pm,i}|i\rangle_{\pm}
\end{equation}
where $|i\rangle_{\pm}$ are the bulk states, dependent on $z$, which form the basis of the effective Hamiltonian. For details about the operators $\mathcal{H}$, its basis,  and more details of how to solve the eigenproblem, see \ref{ap:edge-spectrum-states}.




\begin{figure}[ht]
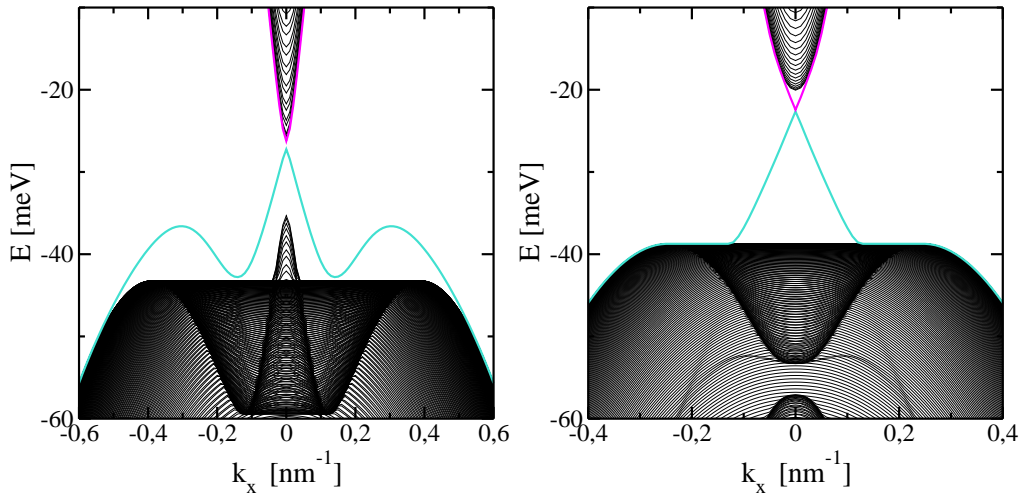

    \centering
    \includegraphics[width=0.4\linewidth]{Fig3a.eps}
    \includegraphics[width=0.4\linewidth]{Fig3b.eps}
    \caption{\label{fig:bulk-edge-spectrum} Spectrum of $\mathbf{H}_{eff}$ as a function of $k_x$, for $d=7$~nm in the left panel, and $d=9$~nm in the right panel. The 
bulk eigenvalues correspond to the black solid lines, while the energy of 
the edge states corresponds to the 
light blue and pink lines. The gap between sub-bands is closed by the edge states for $k=0$.}
\end{figure}

Figure~\ref{fig:bulk-edge-spectrum} shows the spectrum of the effective Hamiltonian as a function of $k_x$ for $d=7$~nm (left panel) and $d=9$~nm (rigth panel). 
The helical edge states can be easily identified, since they lay in the gap between sub-bands, and are represented by the light blue an pink lines. For $d=7$~nm the helical states appear in the gap between the $H1$ and $E1$ sub-bands, and between $H_1$ and $E_2$ for $d=9$~nm. Note that in both cases the system is in the inverted regime, see Figure~\ref{fig:bulk-spectrum}. The spectra in Figure~\ref{fig:bulk-spectrum} are in very good agreement with the ones obtained from the Kane Hamiltonian for small values of $k_x$ \cite{Krishtopenko2018}. Figure~\ref{fig:bulk-edge-spectrum} also shows that, in the absence of an external magnetic field, the spectra are symmetric as functions of $k$, {\em i.e.}, $E_+(+k)=E_-(-k)$. This is not the case when a magnetic field is applied, as will be shown in the next section. Note that the presence of edge states transforms the system, whose energy spectrum is given in Figure~\ref{fig:bulk-spectrum}, in a gapless one. The closure of the gap forms a Dirac's cone, that is a well known signal of the presence of topological states.

\subsection{Two-dimensional varational eigenstates}

Collecting the results from the previous sections, and as a result of some calculations summarized in the Appendixes, we are now in condition to express the two-dimensional edge states of the system.

As stated before, the eigenfunctions of the Kane Hamiltonian, Equation~\ref{eq:kp-ocho-bandas}, that correspond to the $E1$, $E2$, $H1$ and $H2$ sub-bands, for spin up and spin down, are the basis in which the effective Hamiltonian is written. So, each of the components of the eigenfunctions of $\mathcal{H}$ and $\mathcal{H}^*$, $\Phi_{\pm,i}$, is  accompanied by the corresponding sub-band eigefunction. The two-dimensional  variational eigenstates can then be expressed as

\begin{equation}\label{eq:variational-y-z}
|\phi(y,z)\rangle_{\pm} = \sum_{i,k,n,p} \alpha^{i,k,\pm}_{n,p} \psi_n(y) B_p(z)  |k\rangle ,
\end{equation}

\noindent where $i=1,..,4$ and $\pm$ corresponds to the effective Hamiltonian eigenstates, $|k\rangle$ for $k=1,..,8$ is the Kane Hamiltonian basis, $\psi_n$ for $n=1,..,N$ are the functions in Equation~\ref{eq:basis-osc}, and $B_p$ for $p=1,..,M$ are  $B-$spline functions \cite{Bachau2001,deBoor1978}. The coefficient $\alpha^{i,k,\pm}_{n,p}$ contains the coefficients involved in Equations~\ref{eq:variational-eigenfunctions-z}, \ref{eq:En-Hn} and \ref{eq:variational-edge}.



In what follows, we will focus in the characterization of the behaviour of the localized edge states, $|\phi(y,z)\rangle_{\pm}$, for given values of the parameters $B$ and $k_x$, the strength of the magnetic field applied in the $z$ direction and the wave vector in the only direction that does not have finite boundary conditions for the variational eigenstates, respectively. This election for the direction of the magnetic field is not the most general, but corresponds to the most common experimental settings. Also, this choice has several analytical advantages, since  the magnetic field dependent  effective Hamiltonian keeps the block structure of Equation~\ref{eq:effective-Hamiltonian}.

\section{Results: Characterization of states using information-like quantities}\label{sec:entropies}

One of the best known quantum information quantifiers is the von Neumann entropy, which is
given by
\begin{equation}\label{eq:vonneumann-def}
S(\rho_A) = -\mbox{Tr} \rho_A \log_2(\rho_A).
\end{equation}
where $\rho_{A}$, the reduced density operator of a state  $\rho_{AB}$ of a composite system with subsystems $A$ and $B$, is given by
\begin{equation}
\rho_A = \mbox{Tr}_B \rho_{AB}, 
\end{equation}

Diagonalizing the matrix representation of the operator $\rho_A$, the von Neumann entropy can be calculated as

\begin{equation}
 S(\rho_A) = -\sum_{\ell} \lambda_{\ell} \log_2(\lambda_{\ell}), 
\end{equation}

\noindent where $\lambda_{\ell}$ are the eigenvalues of $\rho_A$. 

There are many papers devoted to study the information content of topological-like states in spin chains \cite{Isakov2011,Jiang2012,Grover2013,Lee2013} and in systems with continuum degrees of freedom \cite{Haque2007,Li2008,Zozulya2007,Zozulya2009}. Note that, in spin chains,  the natural way to express the total Hilbert space is as the tensorial product of  subsystems which are given by complementary segments of the chain. The usual procedure to obtain reduced density matrices consists in tracing out a segment of the chain, and the information content is calculated through a given quantum entropy associated to this reduced density matrix.  

The intricate way in which the  variational eigenstates approximations to the helical edge states  depend on  many different variables, see Equation~\ref{eq:variational-y-z}, could lead to a number of possible partitions of the Hilbert space, and each one would result in a different characterisation of the states. In the following, we will study a few of such possibilities, by means of the so called real space entropy \cite{Sterdyniak2012} defined for two dimensional systems. Considering the subsystems $A$ and $B$ as 

\begin{equation}\label{eq:subsystem}
A=\mbox{the whole} \, z \, \mbox{coordinate axis}, \qquad B = \mathcal{D},
\end{equation}
where $\mathcal{D}$ is a domain defined in the $y$ coordinate, the real space entropy is the the von Neumann entropy of the reduced density matrix that result from this election. Tracing out a coordinate to obtain a reduced density matrix as a mean to study 
the properties of the quantum state of a single particle was introduced in 
Reference~\cite{AliCan2005}, and has been used to analyze the behavior of 
Laguerre-Gaussian one-photon states \cite{Giovenale2019} and the binding of 
resonance states \cite{Garagiola2018}, among other applications (see 
Reference~\cite{Sterdyniak2012} and References therein).

For the eigenfunctions $\left|\phi\right\rangle_{\pm}$ (Equation~\ref{eq:variational-y-z}) obtained using the 
variational method, we proceed to calculate a real space  entropy  by tracing out, from the two-coordinate one-particle state 
\begin{equation}
  \rho =  \left|\phi\right\rangle_{\pm \,\pm}\left\langle \phi\right|,
\end{equation}
 a spatial domain 
$\mathcal{D}$ in the $y$ coordinate. From now on, we omit the suffix $\pm$ of the effective Hamiltonian basis for clarity. In this way, we construct a 
one-coordinate reduced density matrix

\begin{equation}\label{eq:rhoD}
\rho_\mathcal{D}(z,z') =  \frac{1}{N} \int_\mathcal{D} \left|\phi 
(y,z)\right\rangle
 \left\langle \phi (y,z')\right| \, dy ,
\end{equation}

\noindent where

\begin{equation}\label{eq:rhod-normalizacion}
N = \int \int_\mathcal{D} \left|\phi (y,z) \right\rangle \left\langle 
\phi (y,z) \right| \, dy \, dz,
\end{equation}

\noindent and $|\phi\rangle$ stands for $|\phi\rangle_+$ or $|\phi\rangle_-$. From now on, $\rho_{\mathcal{D}}$ stands for the reduced density matrix obtained by tracing out the subsystem  $B=\mathcal{D}$.

The eigenvalues $\lambda_{\ell}$ of the reduced density operator in Equation~\ref{eq:rhoD}, needed to calculate the real space entropy of the state, are obtained solving the integral equation
\begin{equation}\label{eq:integral-equation-rhod}
\int \rho_\mathcal{D}(z,z') \varphi_{\ell}(z') \, dz' = \lambda_{\ell} \varphi_{\ell}(z) , 
\end{equation}
where $ \varphi_{\ell}(z)$ are the eigenfuntions of the reduced density matrix. This problem can be cast into an algebraic eigensystem problem, by choosing a set of basis functions and diagonalizing the reduced density matrix. Note that once the subsystem $B$ has been traced out, the remaining density operator depends on the $B-$spline functions and the Kane Hamiltonian basis, so  this two sets of functions are the natural choice for a basis to write the reduced density matrix. The entries of the reduced density matrix are then given by

\begin{equation}\label{eq:rhoD-exp-z}
    [\rho_{\mathcal{D}}]_{ka,lb}=\frac{1}{N} \sum_{i,n,p} \sum_{j,m,q} (\alpha^{i,k,\pm}_{n,p})^*\alpha^{j,l,\pm}_{m,q} I_{nm}^{\mathcal{D}}S_{a,p}S_{q,b},
\end{equation}
where $k$ and $l$ are the indixes for the Kane basis, $a$ and $b$ correspond to the $B-$spline basis,
\begin{equation}\label{eq:rhoD-norm}
    N=\sum_{i,k} \sum_{n,m} \sum_{p,q} (\alpha^{i,k,\pm}_{n,p})^*\alpha^{j,l,\pm}_{m,q} S_{p,q}I_{nm}^{\mathcal{D}},
\end{equation}

\begin{equation}
I^{\mathcal{D}}_{nm} = \int_\mathcal{D} \psi_n^{\star}(y) \psi_m(y) \, dy \, \quad \mbox{and} \quad S_{p,q} = \langle B_p|B_q\rangle .
\end{equation}
 Note that when $\mathcal{D}$ is equal to the whole $y$ axis then
\begin{equation}
    I_{nm}^{\mathcal{D}} = \delta_{n,m},
\end{equation}
where $\delta$ is the Kronecker function.

The eigenfunctions $|\phi\rangle_{\pm}$ are either extended along the width of the quantum well, or strongly localized on its borders. This implies that there is not a natural choice for the domain $\mathcal{D}$, if 
the real space von Neumann entropy were to be used to distinguish between  the 
different states that can be found between the eigenstates of 
Equation~\ref{eq:effective-Hamiltonian}. In what follows we will focus on the results obtained when $\mathcal{D}$ is the whole $y$ axis, while other chosings are deferred to \ref{ap:real-space-entropy}. Also, at the end of this section, we present the results obtained from an alternative entropy that we propose, which further characterize the states of the system.

\begin{figure}[h]
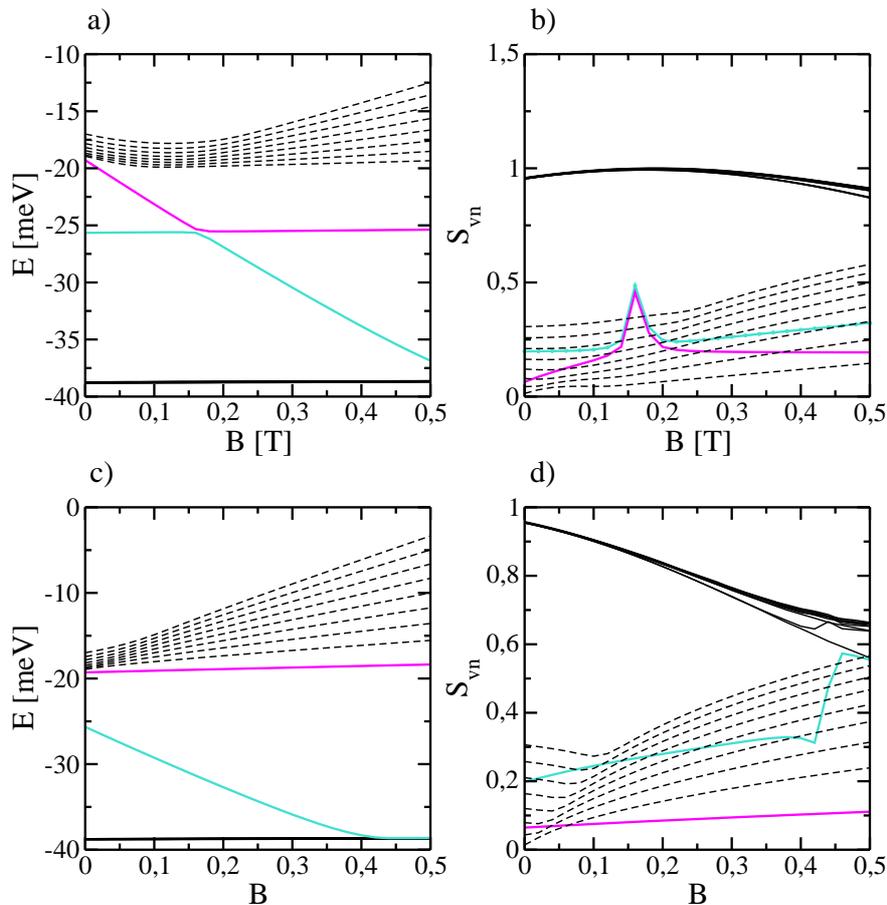

    \centering
    \includegraphics[width=0.7\linewidth]{Fig4a.eps}
    \includegraphics[width=0.7\linewidth]{Fig4b.eps}
    \caption{\label{fig:von-neumann} a) Energy spectrum of the effective Hamiltonian for a quantum well of width $d=9$~nm as a function of the magnetic field intensity $B$, for $k_x=0.02$~nm$^{-1}$. In full black lines are higher energy states of the valence band, the dashed black ones correspond to the lowest energy states of the conduction band, and the pink (highest energy in the gap) and light blue (lowest energy in the gap) lines correspond to the two edge states. b) real space entropy calculated for $\mathcal{D}$ as the whole $y$ axis. The parameters are the same as in a), and the color code corresponds to the same states in both panels. Panels c) and d) have the same information as panels a) and b), respectively, but for $k_x=-0.02$~nm$^{-1}$}
\end{figure}

The results that we present correspond to a very reduced set of parameters ($k$, $d$ and $B$), nevertheless they are representative of the physical traits that are observed for other values of those parameters.  For choosing the quantum well parameters, some considerations must be taken into account. First, from Figure~\ref{fig:bulk-edge-spectrum} it is clear that for very small values of $k$ the helical states are farther away from the bottom and top of the conduction and valence bands, respectively. Also, the various perturbations made to obtain the Kane and effective Hamiltonians, results in accurate solutions for small values of this parameter. Second, when an external magnetic field is applied the helical states eventually enter into the conduction or valence bands. The actual value of the magnetic fields for which the helical states enter the bands depend on $k$, the width $d$, etc. The value of $k$ and the magnitude of the magnetic field are proportionally inverse, so some compromise in choosing the value of $k$ has to be made.

\subsection{Real space - von Neumann entropy for normal and edge states} 

In this subsection, we consider the real space entropy when $\mathcal{D}$ is the whole $y$ coordinate.

Figure~\ref{fig:von-neumann} a) shows the spectrum of the effective Hamiltonian, for a quantum well of width $d=9$~nm, as a
function of the magnetic field strength $B$, for $k_x=0.02$~nm$^{-1}$. The figure shows only several variational eigenvalues of the
conduction and valence bands,  using dashed and continuous black lines, respectively. The helical edge states
energies are shown by the light blue and pink lines, the same color code used in Figure~\ref{fig:bulk-edge-spectrum}. Panel b)
shows the behaviour of the real space entropy for those states included in panel a). Panels c) and d) are equivalent to panels a) and b), respectively, but correspond to $k_x=-0.02$~nm$^{-1}$. The curves in panels b) and d) show an ``inverted ordering'' with respect of panel a) and c), meaning that the valence band states have larger entropies that the states
in the conduction band.

The values of the real space entropy of the states considered, show that the helical states have the same symmetries as the states of the valence band, which are $H1$ states, {\em i.e.}, hole-like states. This behaviour was found for several values of $k_x$ and $d$, since not matter the value of this parameters, the bottom a the valence band is always the $H1$ bulk sub-band. This is not the case for the top of the conduction band, which is the $E2$ sub-band for $d=7,8$~nm and $H2$ for $d=9$~nm \cite{Krishtopenko2018}.

\subsection{Decomposing the Hilbert space in a different way} 

The results obtained concerning the real space entropy shown in Figure~\ref{fig:von-neumann} suggest that other quantity to measure the localization on the $y$ coordinate and the strong polarization of the helical edge states is needed. 

As said before, the elections that can be made of two subsystems for obtaining an information-like entropy are several. Nonetheless, the mathematical features of the edge states obtained using the methods described, leaves a reduced number of options since it is not always possible to write the Hilbert space as a tensor product. An alternative for this treatment is to consider an entropy that is calculated for the hole states, without tracing out a part of the Hilbert space. Many of these quantities were considered in this work, in which we calculated the von Neumann entropy taking into consideration different partitions of the coefficients of the eigenstates obtained.

In this sense, the simpler quantity one can consider is given by
\begin{equation}
    S^{edge}=\sum_n \sum_i |d_n^{i,\pm}|^2\log (|d_n^{i,\pm}|^2)
\end{equation}
where the sum in $n$ is over the eigenbasis in the $y$ direction of, and the one in $i$ is over the basis of the effective Hamiltonian. The coefficients $d_n^{i,\pm}$ corresponds to the solutions of the effective Hamiltonian (see \ref{ap:edge-spectrum-states}).

Note that the $S^{edge}$ entropy is fairly similar to the real space entropy, if the $z$ coordinate were the one traced out.





\begin{figure}[h]
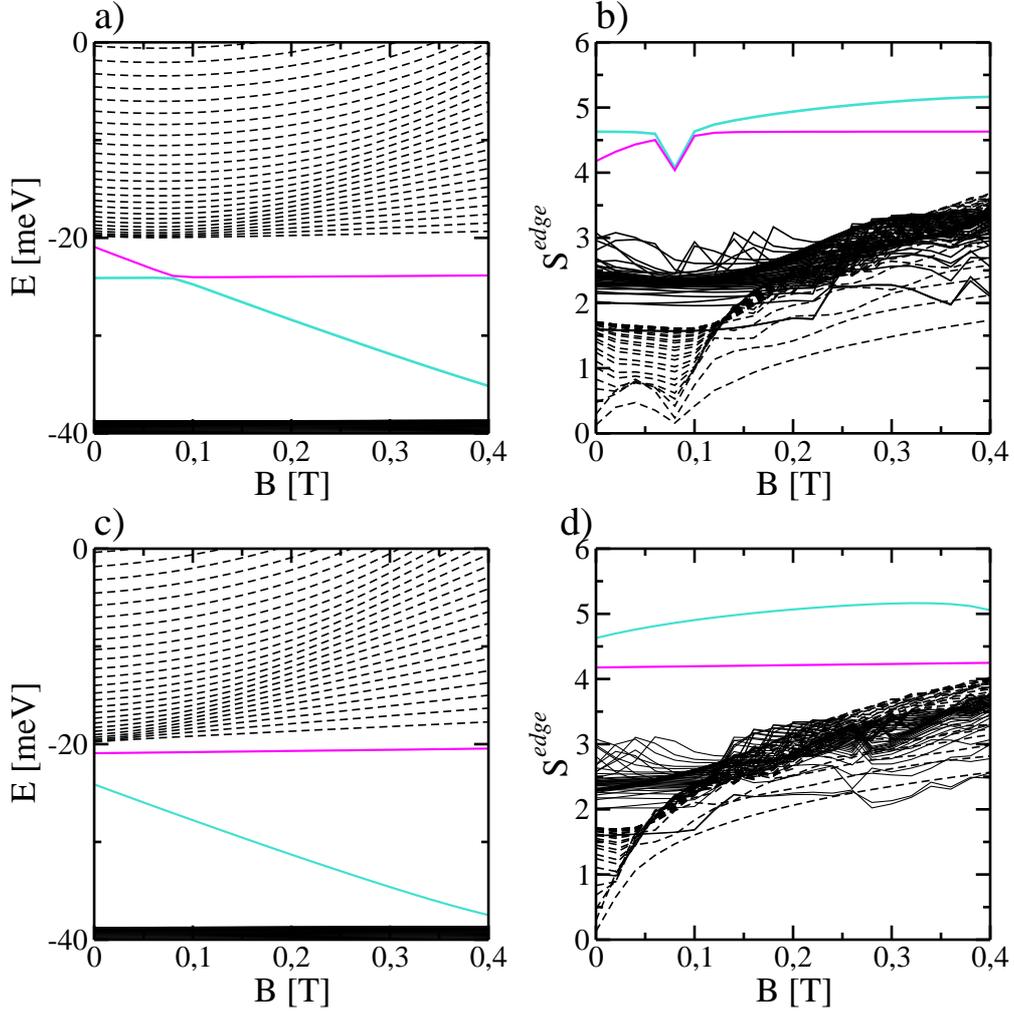

    \centering
    \includegraphics[width=0.8\linewidth]{Fig5a.eps}
    \includegraphics[width=0.8\linewidth]{Fig5b.eps}
    \caption{\label{fig:entropia-clasica} \label{fig:edge} a) Energy spectrum of the effective Hamiltonian for a quantum well of width $d=9$~nm as a function of the magnetic field intensity $B$, for $k_x=0.01$~nm$^{-1}$. The valence band is depicted in black lines, la dashed black lines correspond to the conduction band, and the light blue and pink ones correspond to the edge states energies. b) $S^{edge}$ entropy for the states corresponding to the energies included in a). The color code corresponds to the same states in both panels. Panels c) and d) have the same information as panels a) and b), respectively, but for $k_x=-0.01$~nm$^{-1}$}
\end{figure}

Figure~\ref{fig:entropia-clasica} a) and c) show the energy spectra of a quantum well of width $d=9$~nm as a function of the magnetic field, for $k=0.01$~nm$^{-1}$ and $k=-0.01$~nm$^{-1}$ respectively. The valence band is depicted by black lines, while the conduction band is in dashed black lines. The light blue and pink lines are the helical edge states in this configuration. Panels b) and d) in Figure~\ref{fig:entropia-clasica} show the $S^{edge}$ entropy  for the states in panels a) and c) respectively. The strong polarization of the helical edge states is clearly put into evidence by this measure, since they present entropies grater than any of the bulk states close in energy. A clear differentiation of helical edge states, from normal states, is obtained from this quantity, which virtually does not change for these two states in the parameter interval shown. 

\newpage

\section{Discussion and Conclusions}\label{sec:conclusions}

Even in spite of the simplification that implies to treat the two dimensional 
problem using an effective Hamiltonian, which 
effectively reduces the dimensionality of the numerical algorithms needed to 
find the edge states, the whole procedure is quite taxing and CPU time 
consuming. Nevertheless, we have tested the characterization of edge states 
using different entropies for values of the width $d=7$~nm, $d=8$~nm and $d=9$~nm, 
and 
for a variety of values of $k$ and $B$ for each value of $d$.  In all the analyzed cases, the 
behavior observed was, qualitatively, the same that the one shown in 
Figures~\ref{fig:von-neumann}, \ref{fig:entropia-clasica} and  \ref{fig:ext-km0-29B0-3}. It is clear that different choices for the subspace that is traced out result in a different characterization of the helical edge states. Most notably, $S^{edge}$ clearly identifies the edge states as those that possess more information among the states that are close to the band gap. Even more, this information is almost independent on the strength of the external magnetic field. It would be interesting to study if this trait remains under magnetic impurities.

In both Figures~\ref{fig:von-neumann} and \ref{fig:entropia-clasica} we do not include states that lie too far away from the conduction band bottom or the valence band top. There are at least two reasons to proceed in this way. The first has to do with the continuum nature of the energy  in the bands which makes it difficult to envisage how a given state can be separated from another one that lies close to it in energy. The Rayleigh-Ritz method represents the continuous bands with a discrete set of eigenvalues, whose density is a good approximation to the ideal density of states, but the relationship between the continuum states and the variational ones becomes harder to grasp when states that lie far away from the gap are considered. 

The second reason that supports that the states far from the gap should not be considered comes from the fact that the effective Hamiltonian is obtained using a {\em perturbative} expansion, using the sub-bands states $E1$, $E2$, $H1$ and $H2$.  Into the conduction and valence bands, especially in the later, there is a strong mixing between the sub-bands associated to the sub-bands $E1$ and $H2$, leading to a variational spectrum full of crossings and avoided crossings which make very difficult to follow the state of a single eigenvalue, or even the eigenvalue. This phenomenon is well known and has been pointed in different studies concerning systems with external magnetic fields applied to them where the avoided crossings come from continua associated to different Landau Levels \cite{Garagiola2018,Kishore2014,Ramos2014}.

A closely related subject is the stability of the numerical energy and spinors helical edge states found using the Rayleigh-Ritz method. The values of the energy are rather easily obtained, at least for $k\neq 0$. For $k=0$ it is known that there is no gap between the two helical edge states \cite{Krishtopenko2018}. This feature is particularly difficult to be obtained using the variational method, requiring very large basis sets of $ \psi_n$ functions. Nevertheless, as can be appreciated in Figure~\ref{fig:bulk-edge-spectrum}, for large enough basis sets size ($>2000$) the variational  spectrum is gapless with an error lesser than $10^{-3}$~meV. The other features of the spectrum and entropies are very stable. It is clear that the large basis set sizes are necessary because the helical edge states are localized over lengths that  are much more smaller than the length of the slab in the $y$ direction, nevertheless the value considered is compatible with realistic setups.

As can be seen in Figure~\ref{fig:ext-km0-29B0-3}, the study of the real-space entanglement,
measured 
through the von Neumann entropy, is able to quantify and characterize the localized
behavior of edge states. The behavior of the von Neumann entropy separates 
clearly the localized edge states from the extended bulk ones. Even tough the 
calculus of the different steps necessary to obtain the von Neumann 
entropy is full of  intricacies, it is interesting that the difference between each 
kind of states can be so clearly appreciated.

The domains that we choose to calculate reduced density matrices are not the only possible choices. Choosing disjoint segments $A$ and $B$ is 
one of the more attractive possibilities that come to mind. We are working 
along these lines of research. Other issues worth of further investigation includes the effect of external magnetic fields with components in the $(x,y)$ plane, higher dimensional effective Hamiltonians and the possibility to consider a full two-dimensional treatment of the Kane Hamiltonian.  

\section*{Acknowledgements}

We acknowledge SECYT-UNC and CONICET (PIP-11220150100327CO) for partial 
financial support. We also want to thank Hern\'an L. Calvo for fruitful 
discussions 
about the physics of edge states and Mariano Garagiola for the help with the 
implementation of the numerical algorithms in the first stages of the 
investigation leading to this article.

\appendix

\section{Bulk Espectrum and states}\label{ap:Kane-Hamiltonian}

The matrix elements of the Kane Hamiltonian in Equation~\ref{eq:kp-ocho-bandas} are operators given by

\begin{eqnarray}
\label{eq:kp-ocho-bandas-T}
T &=& E_{c}(z) + \frac{\hbar^2 }{2m_0} \left[
(2F+1) (k_x^2 + k_y^2) + k_z (2F+1) k_z
\right] 
\,,\\
\label{eq:kp-ocho-bandas-U}
U &=& E_{v}(z) - \frac{\hbar^2 }{2m_0} \left[ 
\gamma_1 (k_x^2 + k_y^2) + k_z \gamma_1 k_z \right] 
\,,\\
\label{eq:kp-ocho-bandas-V}
V &=& - \frac{\hbar^2 }{2m_0} \left[ 
\gamma_2 (k_x^2 + k_y^2) - 2 k_z \gamma_2 k_z \right] 
\,, \\
\label{eq:kp-ocho-bandas-R}
R &=&- \frac{\hbar^2 }{2m_0} \frac{\sqrt{3}}{2}\left[ 
(\gamma_3-\gamma_2) k_{+}^2 - (\gamma_3+\gamma_2) k_{-}^2
\right] 
\,,\\
\label{eq:kp-ocho-bandas-Sbar}
\bar{S}_{\pm} &=& - \frac{\hbar^2 }{2m_0} \sqrt{3} k_{\pm} \left[
 \lbrace\gamma_3, k_z\rbrace +   \lbrace \kappa, k_z\rbrace
\right]
,\\
\label{eq:kp-ocho-bandas-Stilde}
\tilde{S}_{\pm} &=& - \frac{\hbar^2 }{2m_0} \sqrt{3} k_{\pm} \left[
 \lbrace\gamma_3, k_z\rbrace - \frac13  \lbrace \kappa, k_z\rbrace
\right] \,,\\
\label{eq:kp-ocho-bandas-C}
C &=& \frac{\hbar^2 }{m_0} k_{-} \left[ \kappa, k_z\right] \, .
\end{eqnarray}
where
\begin{eqnarray}
 \label{eq:kp-ocho-bandas-kpm}
k_{\pm} &=& k_x \pm i k_y\,,\\
k_z &=& -i \partial_z . 
\end{eqnarray}

In Equation~\ref{eq:kp-ocho-bandas-kpm}, $k_x$ and $k_y$ can be either differential operators or vectors in the Brillouin zone, if the quantum wells has boundary conditions in those coordinates or not, respectively. From Equation~\ref{eq:kp-ocho-bandas-T} trough \ref{eq:kp-ocho-bandas-C}, 
all the parameters involved are characteristic of the semiconductor, 
$E_c(z)$ and $E_v(z)$ are the conduction and valence band edges, $\gamma_1, 
\gamma_2$ 
and $\gamma_3$ are the Luttinger parameters, $\Delta$ is the spin-orbit energy, 
$\kappa$ and $F$ take into 
account the interaction with far-lying bands and $P$ is the Kane momentum matrix 
element. The values for the material parameters used are listed in Table~\ref{tab:kane}. As  usual,  $\left[A,B\right]$ and $\lbrace A,B\rbrace$ are the 
commutator and anticommutator of operators $A$ and $B$, respectively. 
\begin{table}[h!]
\begin{center}
\begin{tabular}{|| c c c c c c c c c||} 
\hline
\rule{0pt}{2.5ex} 
  & $E_g(eV)$ & $\Delta(eV)$ & $E_p(eV)$ & $F$ & $\gamma_1$ & $\gamma_2$ & $\gamma_3$ & $\kappa$\\ [0.5ex] 
\hline\hline
\rule{0pt}{2.5ex}  
CdTe & 1.606 & 0.91 & 18.8 & -0.09 & 1.47 & -0.28 & 0.03 & -1.31  \\ [0.5ex] 
\hline
\rule{0pt}{2.5ex}  
HgTe & -0.303 &1.08 & 18.8 & 0 & 4.1 & 0.5 & 1.3 & -0.4  \\ [0.5ex] 
\hline
\end{tabular}
\caption{Parameter in the Kane Hamiltonian por CdTe and HgTe\cite{Novik2005}}
\label{tab:kane}
\end{center}
\end{table}

The eigenvalues shown in Figure~\ref{fig:bulk-spectrum} where obtained 
using a high-precision variational method with $B$-spline functions \cite{Bachau2001,deBoor1978}, $\{B_i\}$, as the 
variational basis set in the $z$ direction. The $B$-spline basis set has a number of properties that 
make them well suited to analyze the one-dimensional 
eigenproblem in Equation~\ref{eq:bulk-eigenproblem}. In particular, they can be 
tuned to 
fulfill the discontinuous derivative of the  functions $\Psi_i$ at the interface 
between different materials accommodating any boundary condition (see Reference \cite{Garagiola2019} and References 
therein). For this reason, although its use is not that common in the context of solving Shcrödinger equations, when compared with the number in which they have been applied in few-body Quantum Mechanics (see References \cite{deBoor1978,Bachau2001} and references therein, there are several works that use them when multidimensional spinors are involved, as are the cases of $\mathbf{k}\cdot\mathbf{p}$ Hamiltonians \cite{Garagiola2019,Giovenale2022} and Dirac Hamiltonians \cite{Grant2009,FillionGourdeau2016}.

Basis sets used for solving eigenproblems are, naturally, finite. This transforms the exact eigenvalue problem into a variational one that can be solve applying the Rayleigh-Ritz method.
In the $B-$spline basis, the eigenstates components for this problem can be expanded as 
\begin{equation}\label{eq:variational-eigenfunctions-z}
\Psi_i(z) = \sum_{p=1}^M c^i_{p} B_p(z), 
\end{equation}
where the index $i=1,2,..,8$ indicates the component correspondent to the $i-$th element of the basis of the Kane Hamiltonian, and $p=1,2,..,M$ is the spline function index. For obtaining the coefficients $c_p^i$, and the variational spectrum associated to these states, the eigenvalue problems in Equation~\ref{eq:bulk-eigenproblem} needed to be solved are now 
\begin{equation}\label{eq:bulk-envelope-variational}
 \sum_{j=1}^{8} \sum_{q=1}^M \mathbf{\bar{H}}_{ip,jq} c^j_{q} = E_B 
 S_{pq} c^i_{q} \; ,
\end{equation}
\noindent for $i=1,..,8$ and $p=1,..,M$, where $\mathbf{\bar{H}}_{ip,jq} = \left\langle B_p | \mathbf{H}_{ij} 
| B_q \right\rangle$, $S_{pq} = \left\langle B_p| B_q\right\rangle$ and $M$ is 
the variational basis set size.

Another advantage in the use of $B-$spline functions in this problem comes from the fact that they are compact support functions. This allows, in a very simple way, to change the density of functions used in some areas of the spatial domain. In this way, in our calculations we took $M=150$, assigning $50$ grid point to the HgTe slab, and $50$ to each of the CdTe slabs, even while the former one is considerably smaller. This was chosen in order to have a better representation of the state that would be localized into the well (HgTe slab) while minimizing numerical effort and also it reflects the discontinuity given by the material interfaces on the spinor components. For more details on $B-$spline functions and the properties mentioned above, see References~\cite{Garagiola2019,Bachau2001,deBoor1978} and References therein.

\section{Edge spectrum and states}\label{ap:edge-spectrum-states}

The explicit form of the $\mathcal{H}$ operators in the effective Hamiltonian in Equation~\ref{eq:effective-Hamiltonian} is given by
  \begin{equation}
 \label{eq:H4x4}
\mathcal{H}(\bf{k})=
\begin{pmatrix}
\epsilon_{\bf{k}} +d_3(\bf{k}) &  -Ak_+ & R_1k_-^2 & S_0k_-\\
-Ak_- & \epsilon_{\bf{k}} -d_3(\bf{k}) & 0 & R_2k_-^2\\
 R_1k_+^2 &  0 & \epsilon_{H2}(\bf{k})& A_2k_+\\
 S_0k_+ &  R_2k_+^2 & A_2k_- & \epsilon_{E2}(\bf{k})
\end{pmatrix}
\end{equation}
where 

\begin{eqnarray}
    {\bf k}&=&(k_x,k_y)   \,, \\
    \epsilon_{{\bf k}}&=&C-D(k_x^2+k_y^2) \,, \\
    \epsilon_{H2}&=&C-M-\Delta_{H1H2}+B_{H2}(k_x^2+k_y^2) \,, \\
    \epsilon_{E2}({\bf k})&=&C+M+\Delta_{E1E2}+B_{E2}(k_x^2+k_y^2)\,,
\end{eqnarray} wehre $\Delta_{H1H2}$ ($\Delta_{E1E2}$) is the gap energy between the $H1$ and $H2$ ($E1$ and $E2$) sub-bands when ${\bf k}=0$, and all of the other parameter involved are given in Table~\ref{tab:eff}. When solving the eigenvalue problem for a finite stripe in the $y$ coordinate, $k_x$ becomes the $x$ coordinate of a vector in the first Brillouin zone, and $k_y=-i\partial_y$.

\begin{table}[h!]
\begin{center}
\begin{tabular}{|| c c c c c c c c ||} 
\hline
\rule{0pt}{2.5ex} 
$d$(nm)  & $C^{\{1\}}$ & $M^{\{1\}}$  & $\Delta_{H1H2}^{\{1\}}$ & $\Delta_{E1E2}^{\{1\}}$ & $B^{\{2\}}$ & $D^{\{2\}}$ &  \\ [0.5ex] 
\hline\hline
\rule{0pt}{2.5ex}  
7 & 1.606 & 0.91  & -0.28 &0.03& 18.8 & -0.09 &  \\ [0.5ex] 
\hline
\rule{0pt}{2.5ex}  
9 & -0.303 &1.08 & 0.5 &1.3& 18.8 & 0 &    \\ [0.5ex] 
\hline\hline
\rule{0pt}{2.5ex} 

\rule{0pt}{2.5ex} 
$d$(nm)   & $R_1^{\{2\}}$  & $R_2^{\{2\}}$  & $B_{H2}^{\{2\}}$   & $B_{E2}^{\{2\}}$   & $A_{2}^{\{3\}}$   & $S_{0}^{\{3\}}$&$A^{\{3\}}$\\ [0.5ex] 
\hline\hline
\rule{0pt}{2.5ex}  
7 & -1006.74 & -43.51 & 711.25 & -29.99 & 336.13 & 44.70 & 1.47\\ [0.5ex] 
\hline
\rule{0pt}{2.5ex}  
9 & -1154.64 & -45.28 & 571.70 & -38.94 & 312.21 & 57.30  & 4.1\\ [0.5ex] 
\hline
\end{tabular}

\caption{Parameters for the effective Hamiltonian \ref{eq:H4x4} for a $HgTe-Cd_{0.7}Hg_{0.3}Te$  quantum well \cite{Krishtopenko2018}. The units of the parameters are: meV for those indexed with $^{\{1\}}$t, meV$\cdot$nm$^2$ for the ones with $^{\{2\}}$, and meV$\cdot$nm for the ones with $^{\{3\}}$.}
\label{tab:eff}
\end{center}
\end{table}

The basis in which this Hamiltonian is expresed are the states $|E1,+\rangle\,|H1,+\rangle\,|H2,-\rangle\,|E2,-\rangle$, and the basis correspondent to $\mathcal{H}^*$ is $|E1,-\rangle\,|H1,-\rangle\,|H2,+\rangle\,|E2,+\rangle$. The eigenstates corresponding to this sub-bands can be obtained from the solutions of the Kane Hamiltonian as following
\begin{eqnarray}\label{eq:En-Hn}
 |En,+\rangle &=& \Psi_1^{(En,+)} |1\rangle +  \Psi_4^{(En,+)} |4\rangle + 
 \Psi_7^{(En,+)} 
|7\rangle, \nonumber \\ 
|En,-\rangle &=&  \Psi_2^{(En,-)} |2\rangle +  \Psi_5^{(En,-)} |5\rangle +  \Psi_8^{(En,-)} 
|8\rangle \nonumber \\ 
|Hn,+\rangle &=&  \Psi_3^{(Hn,+)} |3\rangle \nonumber \\ 
|Hn,-\rangle &=&  \Psi_6^{(Hn,-)} |6\rangle 
\end{eqnarray}

\noindent where $\left\lbrace|k\rangle\right\rbrace_{k=1}^8$ is the basis of 
the Kane Hamiltonian, and the 
functions $|En,+\rangle$ and $|Hn,+\rangle$ are functions with spin up while 
$|En,-\rangle$ and $|Hn,-\rangle$ are functions with spin down. The notation $ \Psi_i^{(E,s)}$ indicates the $i-$th component of the eigenstates correspondent to the energy $E$ and spin $s$ of the sub-band under consideration ($Hn$ or $En$).

The basis set used for solving the variational eigenvalue problem that derives from Equations~\ref{eq:eigen-edgep} and \ref{eq:eigen-edgem} is given by
\begin{equation}\label{eq:basis-osc}
\psi_n(y) = \sqrt{\frac{2}{L}} \sin \left(\frac{n\pi}{L} y\right) ,
\end{equation}
where $L$ is the length of the strip, and $n=1,2..,N$. The value of $N$ depends on the width of the quantum well. We found that the optimal value, comparing with the results from the Kane Hamiltonian, are $N=1950$ for $d=7$~nm, and $N=2500$ for $d=9$~nm. Note that the $\psi_n$ functions satisfy the homogeneous boundary conditions at $y=-L/2$ and $y=L/2$. Each component of the variational eigenfunction of the effective Hamiltonian can then be expressed as
\begin{equation}\label{eq:variational-edge}
    \Phi_{\pm,i}=\sum_{n=1}^N d_n^{i,\pm}\psi_n(y)
\end{equation}

To 
analyze the behavior of the 
edge states when a magnetic field is applied to the quantum well along the $z$ 
direction, both Hamiltonians, $\mathcal{H}(+k)$ and $\mathcal{H}^{\star}(-k)$ 
must be 
modified adding the Zeeman term, and using the Peierless substitution $\partial_i \longrightarrow \partial_i - \frac{e }{\hbar}A_i$. 
\cite{Durnev2016}.

We 
compared our results with the corresponding results obtained previously by 
Krishtopenko {\em et al.} \cite{Krishtopenko2018} and found that both sets of 
values differ in less than $0.005\%$ across the region $k\in \left[ 
-0.5,0.5\right]$~$\mbox{nm}^{-1}$. We also calculated the spectra of bulk and 
edge states for 
quantum well widths of $d=8$~nm and $d=9$~nm, and found that the agreement with 
the
results of Reference~\cite{Krishtopenko2018} was excellent in all the cases. 
These results are not included for the sake of brevity.

\section{Real space entropy} \label{ap:real-space-entropy}

To explore the spatial behaviour of the eigenstates $|\phi\rangle_{\pm}$ into account, we calculate von Neumann entropies for three 
qualitatively different domains, which are

\begin{eqnarray}\label{eq:domains}
 \mathcal{D}_1 &:& y\in \left[ \frac{L}{2} - \Delta, \frac{L}{2} \right] , \nonumber \\
 \mathcal{D}_2&:& y\in \left[ \frac{L}{2} , \frac{L}{2}+ \Delta \right] , \nonumber \\
 \mathcal{D}_1\cup \mathcal{D}_2 &:& y\in \left[ \frac{L}{2}- \Delta , \frac{L}{2}+ \Delta \right] ,
\end{eqnarray}

\noindent where $\Delta \in \left[0, \frac{L}{2}\right]$. So, 
using Equation~\ref{eq:rhoD}, we obtain the reduced density matrices $\rho_{\mathcal{D}_1}$, 
$\rho_{\mathcal{D}_2}$ and $\rho_{\mathcal{D}_1\cup \mathcal{D}_2}$, and then we compute their eigenvalues and the 
corresponding von 
Neumann entropies $S(\rho_{\mathcal{D}_1})$, $S(\rho_{\mathcal{D}_2})$ and $S(\rho_{\mathcal{D}_1 \cup \mathcal{D}_2})$ as 
functions of $\Delta$. Besides, we calculate the quantity $S(\rho_{\mathcal{D}_1})+  
S(\rho_{\mathcal{D}_2}) -S(\rho_{\mathcal{D}_1 \cup \mathcal{D}_2})$, which is called the mutual information \cite{Wolf2008}.

\begin{figure}[hbt]
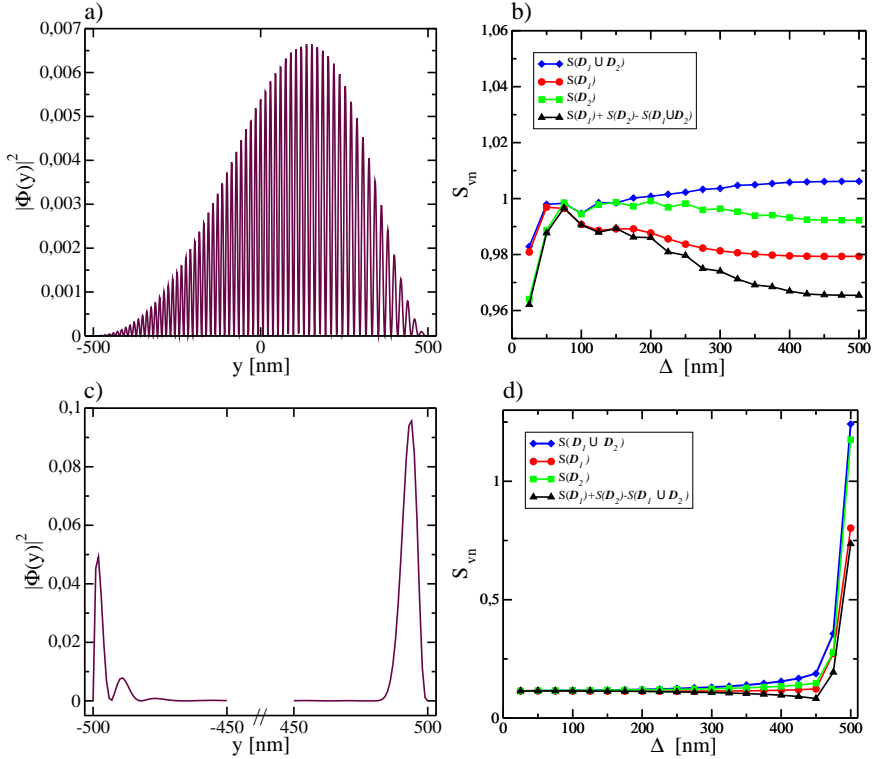

    \centering
    \includegraphics[height=5cm]{Fig6a.eps}
    \includegraphics[height=5cm]{Fig6b.eps}
    \includegraphics[height=5cm]{Fig6c.eps}
    \includegraphics[height=5cm]{Fig6d.eps}
    \caption{\label{fig:ext-km0-29B0-3} The 
Figure shows a) the density probability $|\Phi(y)|^2$ for the state 
immediately bellow the edge state, as a function of $y$, while panel b) 
show the real space von Neumann entropy of the  reduced 
density matrices $\rho_{\mathcal{D}}$, as a function of $\Delta$, for the 
domains $\mathcal{D}_1\cup \mathcal{D}_2$ (blue dots), $\mathcal{D}_1$ (red dots) and  $\mathcal{D}_2$ (greendots), and the quantity $S(\rho_{\mathcal{D}_1})+  S(\rho_{\mathcal{D}_2}) -S(\rho_{\mathcal{D}_1 \cup \mathcal{D}_2})$ (black 
dots). Panels c) and d)  show the same quantities as a), b)  for the edge state. The 
parameters of the system for this Figure are $d=7$~nm, $B=0.3$~T and 
$k_x=-0.29$ $\mbox{nm}^{-1}$. }
\end{figure}

As has been pointed out above, and is shown in 
Figure~\ref{fig:bulk-edge-spectrum}, 
the energy of the degenerate edge state  lies in the gap between two 
sub-bands. This eigenvalue is 
degenerate if no magnetic field, $B$, is applied to the quantum well and the
degeneracy is broken for $B\neq 0$. In what follows we will concentrate on the 
states on the gap and those immediately above and bellow belonging to the 
continuum, {\em i.e.} the top 
and bottom of the sub-bands that are separated by the gap.

The panel a) of Figure~\ref{fig:ext-km0-29B0-3} shows the behavior of 
$|\Phi(y)|^2$ versus the coordinate $y$. Panel 
b) shows the von Neumann entropies 
$S(\rho_{\mathcal{D}_1})$, 
$S(\rho_{\mathcal{D}_2})$ and $S(\rho_{\mathcal{D}_1 \cup \mathcal{D}_2})$ and the quantity $S(\rho_{\mathcal{D}_1})+  
S(\rho_{\mathcal{D}_2}) -S(\rho_{\mathcal{D}_1 \cup \mathcal{D}_2})$ for the state lying at the top of the sub-band 
below the gap, all as functions of $\Delta$. The data shown in 
Figure~\ref{fig:ext-km0-29B0-3} were obtained for $k_x=-0.29$~$\mbox{nm}^{-1}$,  a 
magnetic field 
strength $B=0.3$~T, and for a quantum well with $d=7$~nm and $L=1000$~nm. 
The extended nature of the state in panel a) is clearly manifested 
by $|\Phi(y)|^2$. The corresponding von Neumann entropies (panel 
b)) show a small change when $\Delta$ goes from zero to $L/2$,  note the scale 
of 
the vertical axis. Interestingly the largest changes in  the 
entropies are shown for relatively small values of $\Delta$, around $100$~nm. This behavior is attributable to the effect of the homogeneous boundary 
conditions over the envelope functions. Panels c) and d) of Figure~\ref{fig:ext-km0-29B0-3} show the same quantities that panels a) and b), for the same parameters of the system, but for the only edge 
state present in this configuration. The strong localization of the state near the 
quantum well edges, panel c), results in a very steep behavior of the entropies as can be seen in panel d).

It is worth to mention that the quantity $S(\rho_A)+  S(\rho_B) 
-S(\rho_{A \cup B})$ has been proposed to analyze long range correlations, and topological phases and states \cite{Liu2022}. 
We think that this quantity does not distinguish in a clear way between bulk and edge states.

\bibliography{bib}
\end{document}